\documentstyle[psfig]{l-aa}

\def\bbeta{\mbox{\boldmath$\beta$}}
\def\btheta{\mbox{\boldmath$\theta$}}

\begin{document}
   \thesaurus{12         
              (11.03.4 Abell 2219;  
               12.03.3;  
               12.07.1)} 

\title{Strong and weak lensing analysis of cluster Abell 2219 based on 
optical and near infrared data
\thanks{Based on observations with the William Herschell Telescope
at La Palma, Canary Islands, Spain.}
}

   \author{
J. B\'ezecourt\inst{1} \and 
H. Hoekstra\inst{1} \and
M.E. Gray\inst{2} \and
H.M. AbdelSalam\inst{1} \and
K. Kuijken\inst{1} \and
R.S. Ellis\inst{2, } \inst{3}
   }

   \offprints{J. B\'ezecourt, bezecour@astro.rug.nl}

   \institute{Kapteyn Institute, Postbus 800, 9700 AV Groningen, The Netherlands
   \and Institute of Astronomy, Madingley Road, Cambridge CB3 0HA, UK
   \and Caltech, Pasadena, CA 91125, USA}

\date{Received , Accepted }

\maketitle

\markboth{J. B\'ezecourt et al.: A2219, strong and weak lensing analysis
based on multicolour data}{}

\begin{abstract} 
We present a gravitational lensing study of the massive galaxy cluster
A2219 (redshift 0.22). This investigation is based on multicolour
images from U through H, which allows photometric redshifts to be
estimated for the background sources. The
redshifts provide useful extra information for the lensing models: we
show how they can be used to identify a new multiple-image system (and
rule out an old one), how this information can be used to anchor the
mass model for the cluster, and how the redshifts can be used to
construct optimal samples of background galaxies for a weak lensing
analysis. Combining all results, we obtain the mass distribution in
this cluster from the inner, strong lensing region, out to a radius of
1.5$h_{50}^{-1}Mpc$. The mass profile is consistent with a singular isothermal
model over this radius range.

Parametric and non-parametric reconstructions of the mass distribution
in the cluster are compared. The main features (elongation,
sub-clumps, radial mass profile) are in good agreement.

\end{abstract}

\section{Introduction}
Gravitational lensing is a very powerful phenomenon for determining
the mass distribution in clusters of galaxies at various scales.  In
the inner parts of clusters, giant arcs and multiple image systems
give immediate constraints on the cluster core radius and velocity
dispersion.  Usual values are between 50$h_{50}^{-1}\,kpc$ and
$100h_{50}^{-1}\,kpc$ for the core radius while velocity dispersion
ranges from 800$km\, s^{-1}$ to 1300$km\, s^{-1}$ (Fort and Mellier
1994).  The presence of several multiple image
systems at different redshifts also allows the slope of the mass
distribution with radius to be probed.  Such systems are more easily
discovered with multicolour imaging, including IR, which lead to
photometric estimates of their redshifts, as shown by Pell\'o et
al. (1999a) for two objects at $z=4.05$ in A2390.  Most cluster lens
models show bimodality or elongated structures which are also found in
the X-ray emission in many examples (A370, A2218, A2390, A2104).

In the outer parts, the weak distortions of background galaxies, at a
few percent level, provide a direct mapping of the mass distribution on large 
scales (as reviewed by Mellier 1999). The general shape of the potential is
then accessible as well as substructures or extensions (RXJ1716+67,
Clowe et al. 1998, Hoekstra et al. 2000).

HST images are important for shear measurement thanks to the absence
of a large circularization by seeing. On the other hand, the small
field of view of WFPC2 limits the study to the central parts of the
clusters though in a few cases mosaics of HST images have been
analysed (Hoekstra et al. 1998, Hoekstra et al. 2000). Because of
their angular size, ground based wide field imaging are better adapted
for low redshift clusters.  Mass distributions derived from weak
lensing show a mass to light ratio of several hundred (see the
compilation by Mellier, 1999). Comparison of X-ray and mass maps can
mostly be done for low redshift clusters and it appears that the mass
distribution infered from weak lensing often peaks at the same
location as the X-ray emission (A2218, A370, MS1224+20, MS0302+17).
At high redshift, RXJ1716+67 also shows this agreement. At the same time, 
orientations are quite similar which
means that on large scale the gas traces the mass well.

In this paper we present a combined analysis of gravitational lensing by the 
cluster Abell 2219 and the properties of the lensed sources. We utilise both 
strong and weak lensing constraints in order to determine the mass 
distribution on various scales. A key component of our analysis is the use
of multicolour optical and near-infrared data which is used to 
provide additional constraints on the redshift distribution of lensed sources.

\begin{figure*}
\centerline{\psfig{figure=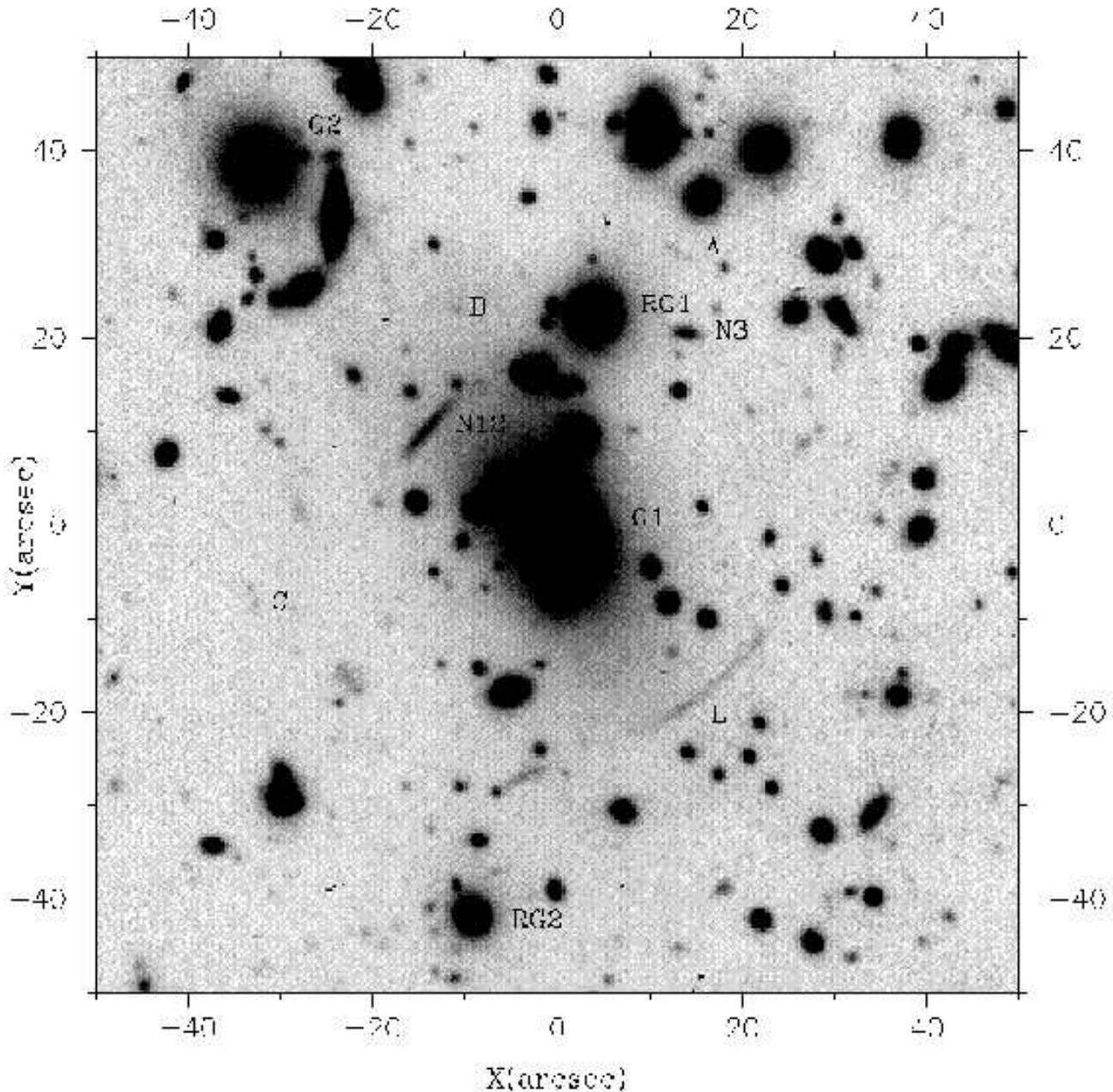,width=\textwidth,angle=0}}
\caption{Composite B and I image of cluster A2219. The objects labels are 
on the right hand side. North is right and east is up.}
\label{cluster}
\end{figure*}

A2219 is a rich cluster at $z=0.225$ (Allen et al. 1992) and is one of
the brightest X-ray clusters detected by the ROSAT All Sky
Survey. Observed also by ROSAT HRI (13.4 ks, Smail et al. 1995) and
ASCA (34 ks, Cagnoni et al. 1998), A2219 has a luminosity of
$L_X(0.1-2.4 keV)=1.8 \times 10^{45} erg\, s^{-1}$ (Smail et al. 1995)
and $L_X(2-10 keV)=3.8 \times 10^{45} erg\, s^{-1}$ corresponding to a
temperature $T_X=9.5keV$ (Allen 1998).  Smail et al. (1995) show the
HRI map to be in a agreement with the general elongated shape of the
cluster on large scale but a misalignement is present on smaller
scales (X-ray emission in A2218, Kneib et al. 1995, doesn't follow the
light either in the very center). The mass derived from lensing
happens to be two times higher than the X-ray mass according to Allen
(1998), though assuming a spherical mass distribution. 
Infrared data at 15$\mu m$ obtained by Barvainis et al. (1999) with ISO
show 5 sources. The 20cm VLA
survey of Abell clusters detected three sources of which the brightest
has a flux of 212 mJy (Owen et al. 1992). These sources are identified
as RG1, RG2 and G2 on figure \ref{cluster}.  Their spectroscopic
follow up identified RG1 as a radio galaxy at $z=0.2070$ (Owen et
al. 1995). Other observations have been done at 28.5 Ghz (Cooray et
al. 1998) and 408 Mhz (Ficarra et al. 1985) for which the peak of
emission coincides with galaxy RG1.

A plan of the paper follows. In section 2 we present the multiband 
observations obtained for this
work as well as already published images. Section 3 is devoted to the
determination of photometric redshifts using 5 filters from $U$ to
$H$. Then, the cluster mass distribution in the central part is
modeled using mainly two systems of multiple images in Section 4. The
weak distortions of background galaxies are studied in the next
section with a comparison of the mass profiles coming from these
various methods.

\section{Observations}
Observations of cluster A2219 have been conducted at the 4.2m William
Herschell Telescope at La Palma, Spain. Images in $B$ and $I$ band
were acquired at prime focus in may 1998 in excellent conditions with
a seeing of 0.8\arcsec \, in $B$ and in $I$. Exposure time is 2400s in
$B$ and 3600s in $I$. In each filter, two pointings of the
2K$\times$4K camera resulted in a wide field of
16\arcmin$\times$15\arcmin \, with 0.237\arcsec/pixel centered on the
cluster.

The infrared observations were made over three nights in June 1998 at WHT 
in june 1998 with the Cambridge Infra Red Survey Instrument (CIRSI). The
individual exposures were either 30 or 60 seconds long, and totalled 2.48
hours. CIRSI is made of four 1k$\times$1k
chips with 0.32\arcsec/pixel. After shifting and adding the images,
the useful field of view in the cluster center is
4.8\arcmin$\times$4.8\arcmin.

In order to cover a wavelength range as broad as possible, we use also
the $U$ image obtained by Smail et al. (1995) at the 5m Hale telescope
at Palomar. Finally, their $V$ image taken at Keck complements this
multicolour description of cluster A2219.

Flux calibration was made with standard stars in M92 (WHT Prime Focus
user manual) and from Landolt (1992). Image reduction was performed
using IRAF packages.  The $1\sigma$ limiting surface brightness for
each image is, in $mag/\arcsec^2$: $U=26.3$, $B=27.2$, $V=27.3$,
$I=25.9$ and $H=23.6$.

\section{Photometric estimates of the redshifts}
The acquisition of images in five filters from $U$ to $H$ enables us
to derive photometric redshifts for the inner part of cluster
A2219. Object detection is performed in the $I$ band with the
SExtractor package (Bertin and Arnouts 1996) with the requirements of
a minimal area of 5 pixels and a detection threshold of 1.5 $\sigma$.
The resulting catalog contains 1427 objects in a
4.8\arcmin$\times$4.8\arcmin square.

Photometric redshifts are computed using the code {\it hyperz}
(Pell\'o et al. 1999b, Bolzonella et al. 2000 in preparation) 
with the following prescriptions:

-- four star formation histories are
adopted: an instantanneous burst, exponentially decreasing star
formation rates with characteristic times scales of 1 Gyr and 5 Gyr,
and a constant star formation. 

-- four metallicty abundances
are considered: $Z_{\odot}/50, Z_{\odot}/5, Z_{\odot}$ and $5
Z_{\odot}$. 

-- internal extinction was allowed to vary in the range $A_V=0$ to 1.2 mag.

\subsection{Redshift distribution of the whole sample}
\begin{figure}
\centerline{\psfig{figure=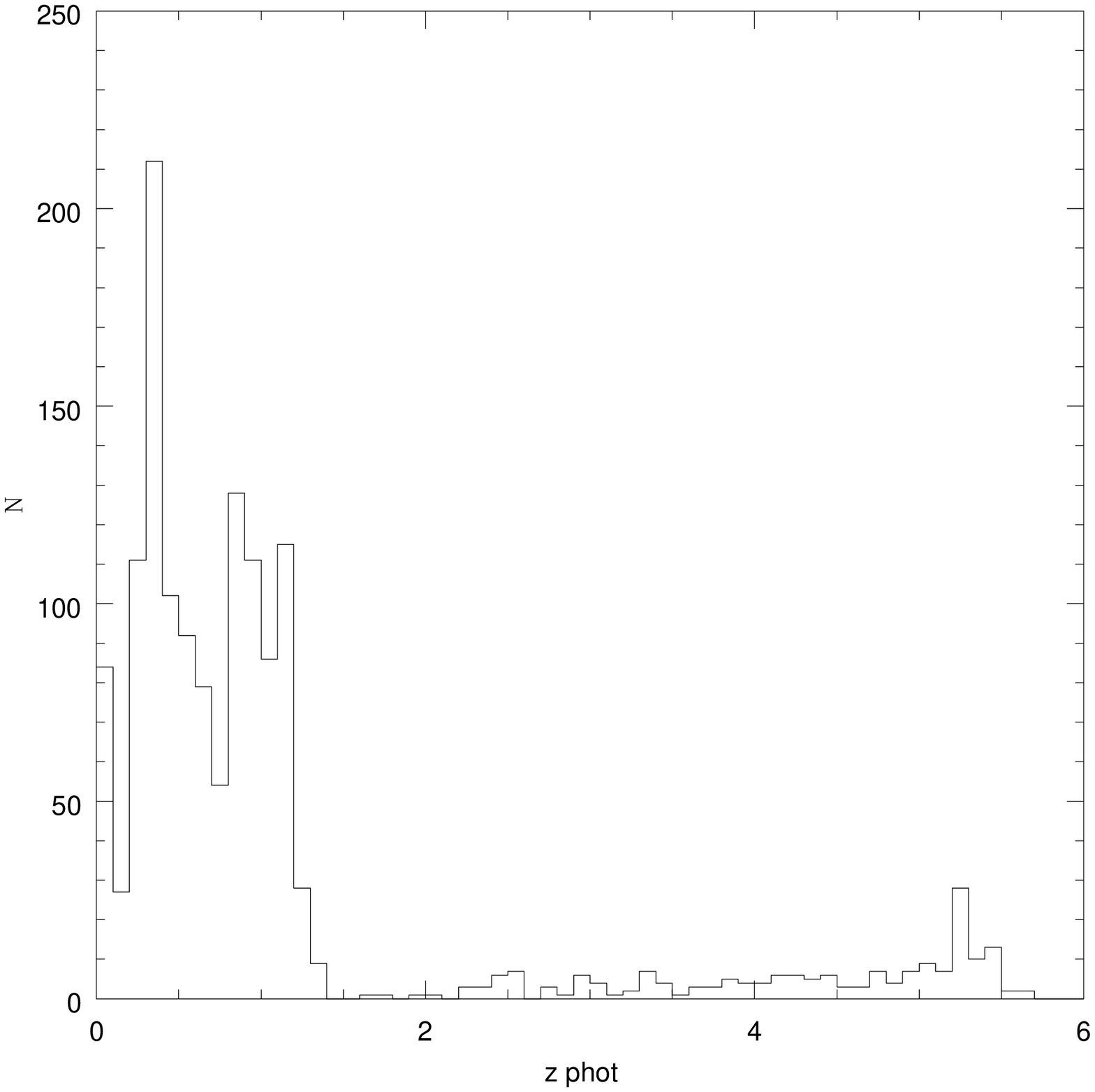,width=0.4\textwidth,angle=0}}
\caption{Photometric redshift distribution in the field of A2219 
derived with filters $UBVIH$.}
\label{N_z}
\end{figure}

\begin{figure}
\centerline{\psfig{figure=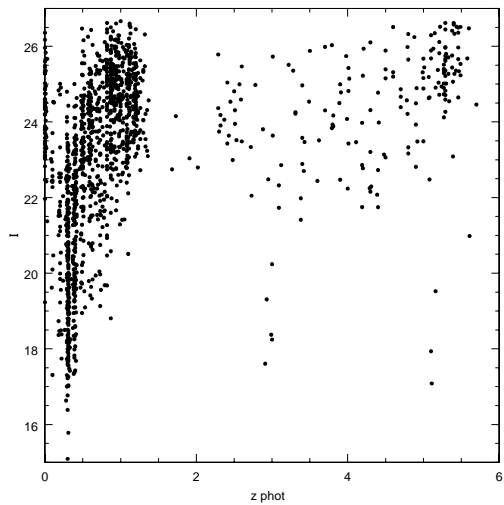,width=0.4\textwidth,angle=0}}
\caption{I - $z_{phot}$ relation for the 1427 objects 
detected in $I$ in the central part of the cluster. 
}
\label{mag_z}
\end{figure}

The global redshift distribution of galaxies in the cluster center is
displayed in figure \ref{N_z}. The peak of the distribution appears
at $z=0.35$ which compares well with the cluster redshift
($z=0.225$, difference of $\simeq 0.1$ in redshift). The second peak 
around z=1 may be an artefact caused by
the absence of any $R$ image.  The 4000\AA\ discontinuity redshifted
at $z=1$ lies between the $V$ and $I$ filters hence $R$ band
photometry would locate the objects redshift with a much better
accuracy when the 4000\AA\ break lies in this wavelength range.
Photometric redshifts higher than 3 still require individual
inspection as most of them correspond to faint objects which are only
detected in the $I$ or $V$ band (figure \ref{mag_z}).

\begin{figure}
\centerline{\psfig{figure=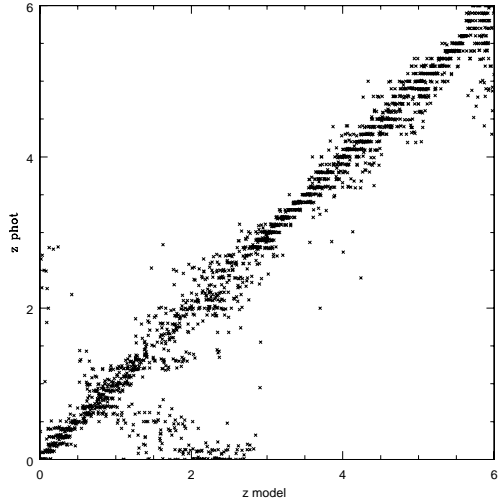,width=0.4\textwidth,angle=0}}
\caption{Photometric redshift versus model redshift for 2000 galaxies
using simulated data in $U, B, V, I,$ and $H$.}
\label{simulation}
\end{figure}

The reliability of these photometric redshifts can be estimated through
simulated catalogs. 2000 galaxies were simulated with magnitudes in $U, 
B, V, I, H$ and randomly distributed from
$z=0$ to $z=6$, with the same metallicities, star formation 
rates and absorption as before.  
Figure \ref{simulation} shows their photometric redshifts derived 
by the Pell\'o et al. code versus the input model redshifts.
Some anomalies in Figure \ref{N_z} can be explained via these simulations.
It appears that the secondary peak in the redshift distribution at $z\simeq 1$
may be an artifact caused by the degeneracy in the method and the
cluster peak appears shifted to a slightly higher redshift for a similar 
reason. No great significance should be attached to sources with $z>3$. 
The simulation was considering a typical
photometric error of 0.1 mag. while the objects with $z_{phot}>3$ in 
A2219 have much higher uncertainties which can reach 0.5 or 1 mag. 

\begin{figure}
\centerline{\psfig{figure=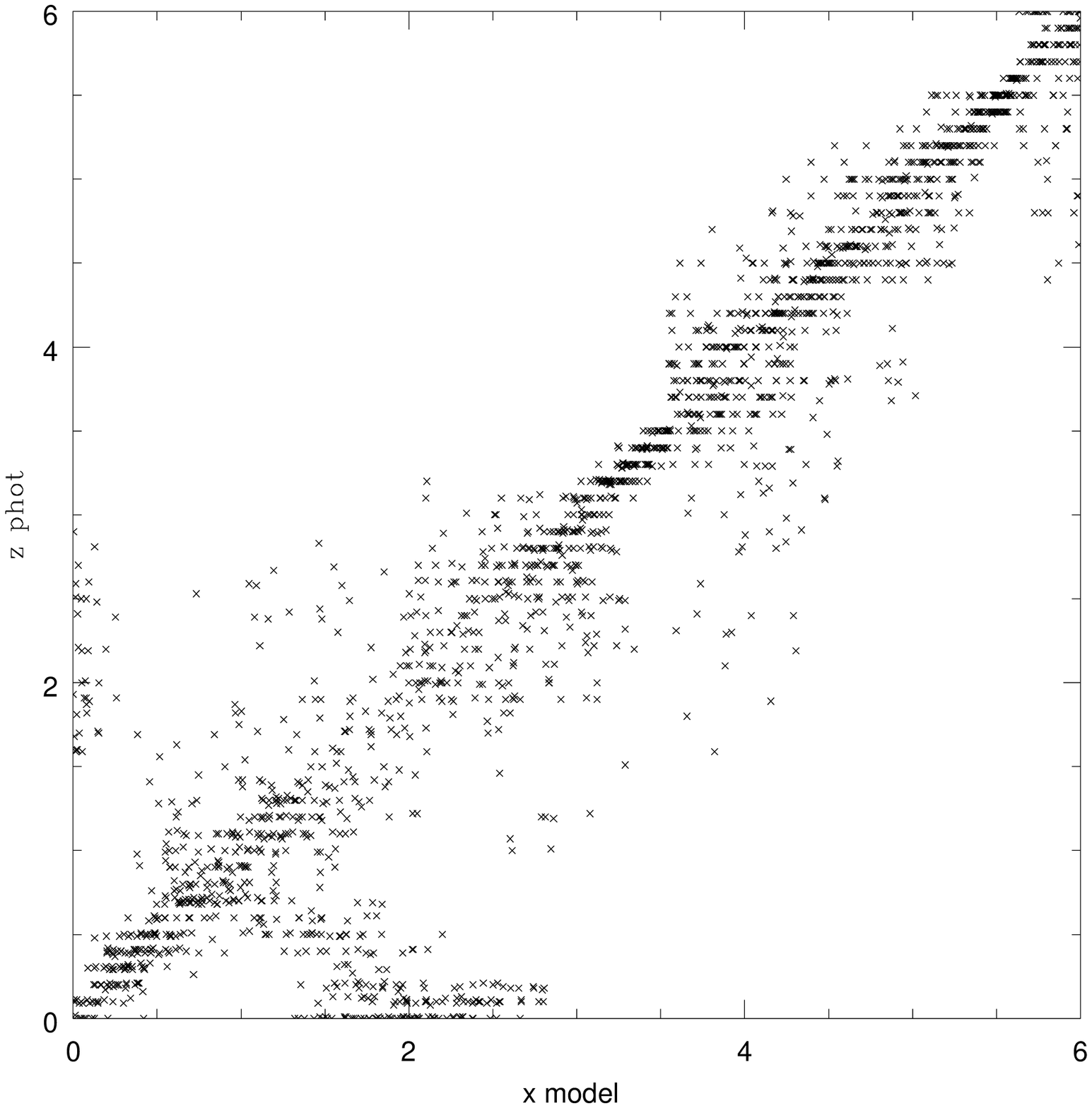,width=0.4\textwidth,angle=0}}
\caption{Photometric redshift versus model redshift for 2000 galaxies
using simulated data in $U, B, V$ and $I$.}
\label{simulation_UBVI}
\end{figure}

The best illustration of the importance of an $H$ band image is given by 
comparison of the simulations in figures \ref{simulation} and 
\ref{simulation_UBVI}.
The first one, computed using information in the IR, shows a
much better correspondance between the model and the photometric redshifts
than the second one which doesn't make use of any measurement in the IR.
In particular, the redshift determination around $z\simeq 2.5$ is worse 
without data in $H$ as the 4000\AA\ break should lie between $I$ and the 
missing $H$. 

\subsection{Individual cases: the triple arc and a new high redshift
multiple system} 
The technique of using multicolour data to identify multiple images
systems in clusters lenses has been demonstrated as a very powerful
technique by Pell\'o et al. (1999a). Two such examples at $z=4.05$ in
cluster A2390 have been studied in details, giving some clues on their
stellar content.

\begin{figure}
\centerline{\psfig{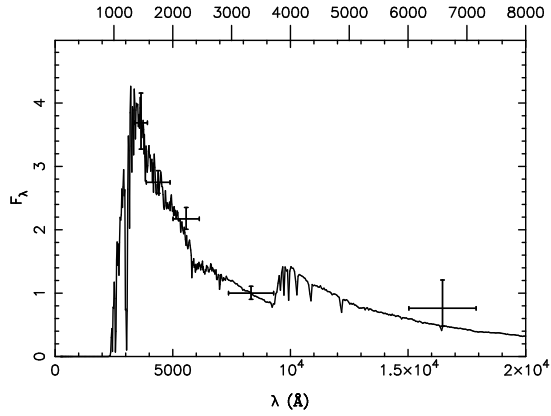}}
\caption{Spectral energy distribution of the triple arc L. The point in $H$ is
an upper limit. The solid line corresponds
to a 0.09 Gyr burst of star formation with solar metallicity at $z=1.5$.
Wavelength in rest frame (\AA) is given at the top.}
\label{sed_L}
\end{figure}

\begin{figure}
\centerline{\psfig{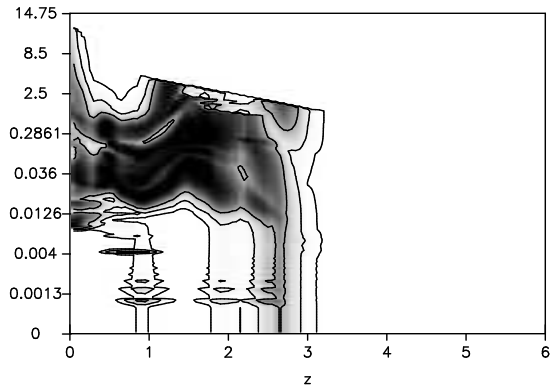}}
\caption{Probability map for the solutions in (redshift, age) for the triple 
arc L. The solid lines are the contours at 68\%, 95\% and 99\% confidence level.
Age is given in Gyr.}
\label{param_L}
\end{figure}

In A2219, Smail et al. (1995) mentioned a very blue giant arc (object L)
consisting of three images.  All three images have the same colour and
none of them is detected in the $H$ band. The arc SED reveals a
quickly rising spectrum towards the UV (figure \ref{sed_L}). However,
the redshift is poorly constrained by photometric data as no strong
discontinuity is enclosed by any filters.  The acceptable domain in
the parameter space (redshift, age) is highly degenerate extending
from $z=0$ to $z=2.6$ with an age between 0.02Gyr and 2.5 Gyr (figure
\ref{param_L}).

\begin{figure}
\centerline{\psfig{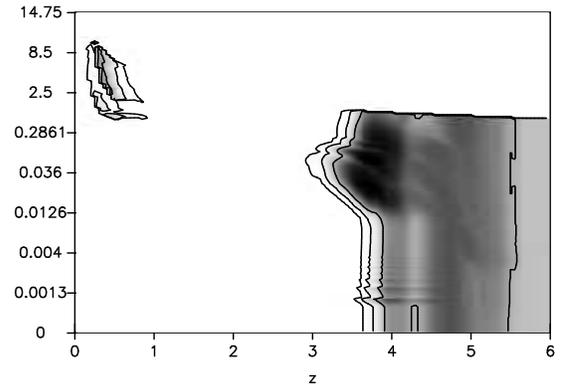}}
\caption{Probability map for the solutions in (redshift, age) for object A.
The solid lines are the contours at 68\%, 95\% and 99\% confidence level.
Age is given in Gyr.}
\label{param_3.6}
\end{figure}

\begin{figure}
\centerline{\psfig{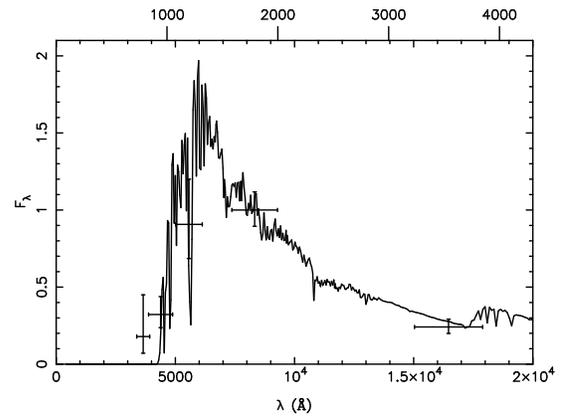}}
\caption{Spectral energy distribution of object A. The point in $U$ is
an upper limit. The solid line corresponds to a burst of star formation seen at 
an age of 0.03Gyr with 5$Z_{\odot}$ at $z=3.6$.}
\label{sed_3.6}
\end{figure}

A second multiple image system seems to be present in A2219 at much
higher redshift. Three red objects appear in the cluster center,
namely objects A, B and C in figure \ref{cluster}. Objects A and B
show very similar SEDs, undetected in $U$ and very faint in $B$ and
$H$.  Object C is even fainter but contaminated by a neighbouring blue
object.  A good solution for object A appears at $z=3.6 \pm 0.4$ with
an age ranging from 0.01Gyr to 1Gyr (figure \ref{param_3.6}).  Object
B is in good agreement with this result in spite of contamination by
the cD red enveloppe, and object C to a lesser extent since its magnitude
measurement is more uncertain due to its faintness and possible
contamination in the case of the $B$ image. Another solution at low redshift
appears to be possible but with a much smaller significance as the
expected IR flux should be much higher than what is
observed. Moreover, the elongated shape and the position angle of A
are consistent with what is expected for a lensed object and B shows
no distortion which is also understandable as it lies in an area with
a smaller shear. Hence, we adopt a photometric redshift of $3.6\pm0.4$
for A and B (figure \ref{sed_3.6}).

\begin{table*}
\caption[]{Photometry of lensed objects in A2219 according to the numbering
by Smail et al. (1995). 
Coordinates are given with respect to the central cD. Object C is located 
close to a blue object and object B is probably contaminated by the red cD 
enveloppe.}
\label{table_arcs}
\begin{flushleft}
\begin{tabular}{ccccccccccccc}
\hline\noalign{\smallskip}
Object & X ($''$) & Y ($''$) &$U-B$&$\sigma_{U-B}$&$B$&$\sigma_B$&
$B-V$&$\sigma_{B-V}$&$B-I$&$\sigma_{B-I}$&$B-H$&$\sigma_{B-H}$ \\
\noalign{\smallskip}
\hline\noalign{\smallskip}
\noalign{\smallskip}
A & 12.9 & 29.6 &-&-&27.00&0.33& 1.74& 0.45&3.16&0.35&4.02&0.39 \\
B & -10.4 & 23.4 &-&-&27.41&0.80&2.15&0.87&3.48&0.84&4.48&1.28 \\
C & -32.5& -8.3 &-& -& 27.07&0.25&0.82&0.39 &2.39&0.32& - & - \\
N$_{1+2}$&-13.7&10.7 &-0.77&0.13&22.65&0.08&0.65&0.15&1.82&
0.11&3.08&0.30\\
N$_3$&14.6 &21.0 &-0.52&0.22&23.49&0.14&0.49&0.23&2.18&0.19&
3.99&0.27\\
L$_3$&-3.6 & -26.5&-0.73&0.13& 23.31&0.05 &0.20&0.09 &0.83&0.11 &-&-\\  
\noalign{\smallskip}
\hline
\end{tabular}
\end{flushleft}
\end{table*}

\section{Mass modeling of the inner part of the cluster}
A2219 is a luminous X ray cluster, one of the brightest clusters seen
in the ROSAT All Sky Survey (Allen at al. 1992).  The general shape of
its X-ray emission obtained with ROSAT HRI shows an elliptical
morphology aligned with the axis defined by the two brightest galaxies
with a luminosity of $L_X=1.8 \times 10^{45}$ erg s$^-1$.  (Smail et
al. 1995). A first mass model was derived by these authors based on
two multiple image systems: the triple arc L and objects $N_{1+2}$ -
$N_3$. However, it seems that N$_3$ is not likely to be a counter image
of N$_{1+2}$ as their colour indices $B-I$ and $B-H$ are clearly
different (table \ref{table_arcs}). Moreover, the redshifts of objects 
$N_{1+2}$ and $N_3$ are
poorly constrained by their photometry. Hence, we have only used the two
systems already described, the arc L and the faint red objects A and B.
Investigation of the mass distribution in the cluster center was
made in two independent ways following methods developed by Kneib et
al. (1996) and AbdelSalam et al. (1998a).

\subsection{Cluster mass distribution as a superposition of a cluster 
and a galaxy components} 
The Lenstool facility developed by J.P. Kneib allows us to consider the
cluster mass distribution of A2219 as the superposition of potentials
centered on the two brightest galaxies and potentials on all galaxies
brighter than $I=19$. All potentials follow a truncated pseudo
isothermal elliptical mass distribution (Kassiola and Kovner 1993).

Following Kneib et al. (1996), for each galaxy halo, the velocity
dispersion $\sigma_0$, the truncation radius $r_t$ and the core radius
$r_0$ are scaled to the galaxy luminosity computed from the observed
$I$ magnitude.  The scaling relations used for the galaxy halos are:
\begin{equation}
\sigma_0=\sigma_{0\ast} \left({L\over L_{\ast}}\right)^{{0.25}},
\end{equation}
\begin{equation}
r_t=r_{t\ast} \left({L\over L_{\ast}}\right)^{{0.5}},
\end{equation}
\begin{equation}
r_0=r_{0\ast} \left({L\over L_{\ast}}\right)^{{0.5}}.
\end{equation}
The scaling relations adopted are motivated by the properties of the 
Faber Jackson relation. 

The orientation and ellipticity of the galaxy halos are taken from the
observed values of the light distribution while $\sigma_{0\ast}$
corresponds to a mass to light ratio of 10 in $I$.  $r_{0\ast}$ and
$r_{t\ast}$ are fixed at 0.15 and 20 $h_{50}^{-1}\,kpc$.


The cluster components are modeled by two large scale mass
distributions centered on the two brightest galaxies.  Their
orientations, ellipticities, velocity dispersions, core radius and
truncation radius are left as free parameters.

\begin{figure}
\centerline{\psfig{figure=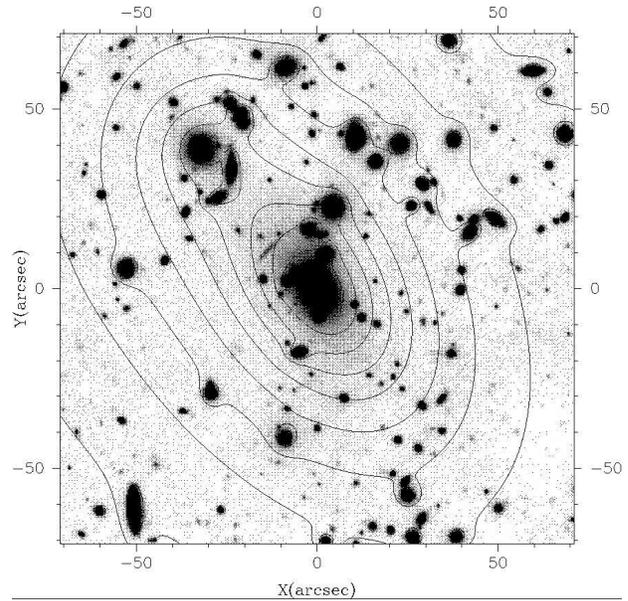,width=0.45\textwidth,angle=0}}
\caption{Mass distribution in A2219 represented by a cluster component
plus a galaxy component overlayed on the $I$ band image.}
\label{mass_jpk}
\end{figure}

The optimisation was made using the constraint that objects A and B
are two images of the same source at $z=3.6$ and that arc L is a
triple arc.  The resulting value for the arc redshift is $z_L\simeq
1.5$ which lies inside the large allowed range derived by the
photometric analysis.  A third image for the $z=3.6$ system is
predicted to be at $x=-36$\arcsec and $y=-3$\arcsec, 6\arcsec\, away 
from object C. The cluster mass
distribution is shown in figure \ref{mass_jpk}.  The main clump has a
velocity dispersion of 1120 $km\, s^{-1}$ and the second one has
540$km\, s^{-1}$.  The cluster mass inside a radius of 150 $h_{50}^{-1}\,kpc$ is
1.2 10$^{14} M_{\odot}$, similar to AC114 (Natarajan et al.  1998),
and $M=3.4\, 10^{14}\, M_{\odot}$ inside 300$h_{50}^{-1}\,kpc$, which is 30\%
smaller than in A370 (B\'ezecourt et al. 1999).

\subsection{Non-parametric mass reconstruction}

The non-parametric reconstruction method works with a pixellated mass
distribution in the lens plane, say $N\times N$ pixels with
inter-pixel spacing $r$. Consider the $mn$-th pixel is a Gaussian
circle with dispersion $r/2$ and peak height $\kappa_{mn}$ in units of
the critical surface mass density. For a background source at an
unlensed position $\bbeta$, the appropriately scaled time-delay in the
direction $\btheta$ is
\begin{equation}
\tau(\btheta)=\frac{1}{2}(\btheta-\bbeta)^2-\frac{D_{ds}}{D_s}\sum_{mn}\kappa_{mn}\psi_{mn}(\btheta),
\label{tau}
\end{equation}
where
\begin{equation}
\psi_{mn}(\btheta)=\frac{1}{\pi}\int_{mn} 
e^{\frac{-2\xi^2}{r^2}}
\ln |\xi| d^2\theta',
\label{psi}
\end{equation}
where $\xi=\btheta^\prime-\btheta_{mn}$. The quantity $\psi_{mn}$ is
the coefficient of the deflection potential at $\btheta$ due to the
$mn$-th pixel only. Thus the contribution of the $mn$-th pixel to the
lens potential is $\kappa_{mn}\psi_{mn}$ (AbdelSalam et al. 1998a).

Lensing observations in clusters of galaxies are:\\ 
$\bullet$ Positions of multiple images on the sky, i.e
$\nabla\tau(\btheta)=0$.\\ 
$\bullet$ Orientations and elongation of individual faint distorted
objects, for example magnification in direction $\theta_x'$ is at
least $\delta$ times that along perpendicular direction $\theta_y'$,
then 
\begin{equation}
\delta|\frac{\partial^2}{\partial\theta_{x'}^2}\tau(\btheta)|\le
  |\frac{\partial^2}{\partial\theta_{y'}^2}\tau(\btheta)|.
\label{ineq}
\end{equation}

Both observations provide us with linear constraint equations on the
unknowns $\bbeta$ and $\kappa_{mn}$. Using constraints from both above
observations combines the strong and weak lensing regimes
simultaneously and breaks the mass-sheet degeneracy upon using at least
two different source redshifts (AbdelSalam et al 1998b). 

Mass maps are then reconstructed by use of quadratic programming to
minimize $M/L$ variations while satisfying the lensing constraints
exactly. So we minimize
\begin{equation}
 \sum_{mn}\left[\kappa_{mn}-L_{mn}\sum_{i'j'}\kappa_{i'j'}\right]^2+\epsilon^4\sum_{mn}(\nabla^2\kappa_{mn})^2,
\label{minmize}
\end{equation}

where $L_{mn}$ is the light distribution of the cluster and $\epsilon$
is a smoothing parameter. 

The mass distribution was determined according to the same 
constraints as before plus the orientation and ellipticity of the faint 
distorted objects ($\sim$ 20 points on a grid inside the $143\arcsec$ field 
excluding the strong lensing region). We assumed a redshift of 1 for 
these objects. In the modelling, two main clumps appear centered on the two
brightest galaxies with offsets of about 2.5 \arcsec\, for G1 towards
the direction of G2 and about 6\arcsec\, for G2 towards G1. As a
smoothing by a Gaussian with $\sigma=3.39\arcsec$ was applied, the
mass peaks coincides well with the light peaks.  Moreover, an
extension of the mass distribution towards the upper right of the cD
agrees well with the presence of several galaxies in this part of the
cluster.

On reconstructing the multiple image system at $z=3.6$ given only
constraints from objects A and B, we find that they are the outcome
of a seven-image system configuration. The image C (which was not included
in the input constraints) is predicted exactly as where it is in the
cluster (see figure \ref{cluster}). For the four remaining images,
two lie in the cD and two are located in empty places.

\begin{figure}
\centerline{\psfig{figure=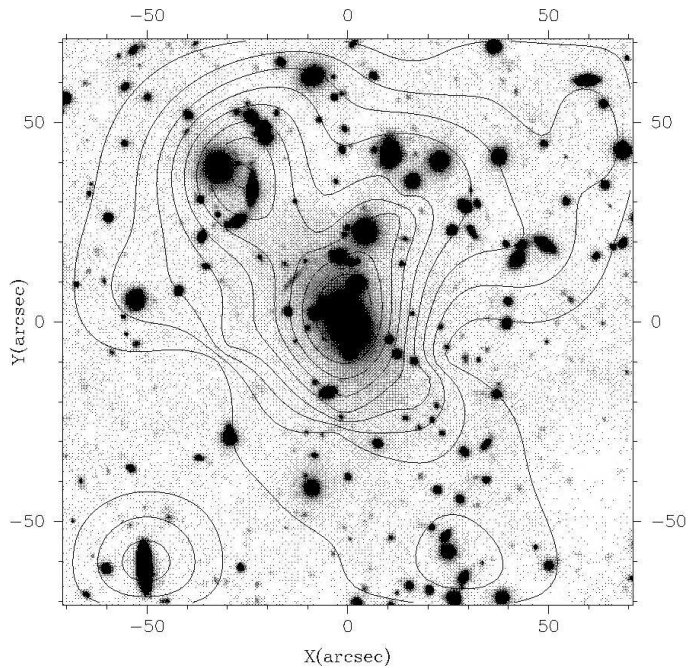,width=0.5\textwidth}}
\caption{Mass distribution in A2219 derived by AbdelSalam et al. (1998a) 
method. The shear field measured in section 5 is indicated.}
\label{mass_hanadi}
\end{figure}


The mass profiles from both methods (Kneib and AbdelSalam et al.) 
for the cluster center agree well
within 25\% at the location of the multiple image systems (figure \ref{mass})
and increase at the same rate with radius. 
The ratio of the main clump, centered on the cD and within 20\arcsec,
to the second one, centered on galaxy G2 and within 20\arcsec, is 1.32
and 1.39 according to the first and second method respectively which show
an excellent agreement. Both models require a bimodal mass distribution 
with a possible extension to the north west.
The triple arc is well reproduced by the models, as well as objects
A and B but object C or possible additional images remain to be investigated.

\section{Weak lensing analysis}

At large radii from the cluster centre, the distortion in the shapes of the
background galaxies is small. This is the regime of weak gravitational
lensing. The lensing signal is obtained statistically, by averaging the 
shapes of 
many background sources. However, the observed shapes cannot be used directly,
as various observational effects, like PSF anisotropy and seeing, have changed
the images of the galaxies.

To measure the weak distortion we follow the procedure described in Kaiser 
et al. (1995), Luppino \& Kaiser (1997), and Hoekstra et al. (1998).

\subsection{Objects catalogs}

The first step in the analysis is to detect the images of the faint galaxies.
Object detection was done in each individual image in $B$ and $I$ with
the {\tt imcat} software (Kaiser et al. 1995), requiring a significance higher
than $4 \sigma$ over the local sky background. The detected objects are
not required to have a photometric redshift estimated.

We use images that were formed by combining the individual exposures by 
straight averaging 
(this avoids corrupting the PSF). Therefore cosmic rays are still present 
in the images.
In the object catalogs we identify very significant objects, but smaller than 2 pixels, 
as cosmic rays. These are removed from the object catalogs. 

Then the objects are analyzed, and sizes, magnitudes, and shape parameters are
estimated. We remove objects for which the analysis failed. From both the $B$ 
and $I$ images this results in a catalog of 13207 objects, of which 5071 are 
detected
in both the $B$ and $I$ band. Restricting the sample to objects lying outside
the cluster elliptical sequence (defined by $2.4<B-I<3.2$) amounts to
11999 and 3863 objects respectively. Star-galaxy separation was done by
plotting the apparent magnitude versus half-light radius. This allows us
to select moderately bright stars to study the PSF anisotropy. At faint
magnitudes stars and galaxies cannot be separated, but at these levels the
galaxies dominate the counts.

\subsection{Shape measurements and corrections}

Objects shapes are characterized by the polarization with its
two components $e_1$ and $e_2$. The polarization is a combination
of the second moments $I_{ij}$:

$$e_1={I_{11}-I_{22} \over I_{11}+I_{22}}$$

$$e_2={2I_{12} \over I_{11}+I_{22}}$$

where $I_{ij}=\int W(\theta)\theta_i \theta_j f(\theta) d^2\theta$,
$f(\theta)$ is the surface brightness and $W(\theta)$ is a Gaussian weight 
function.

\begin{figure}
\centerline{\psfig{figure=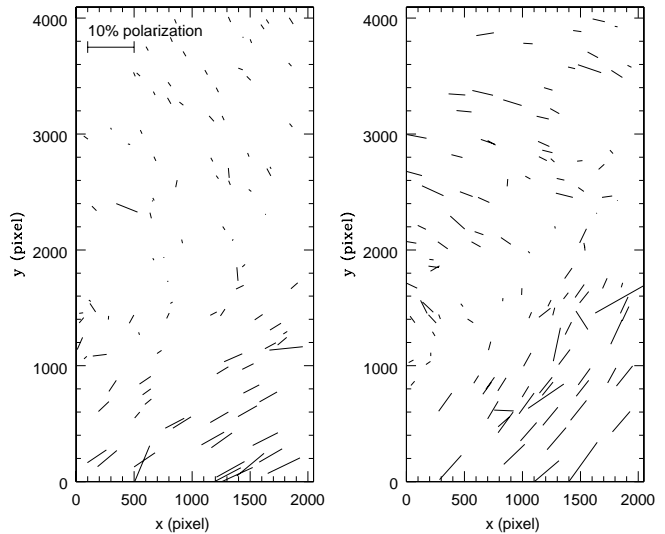,width=0.5\textwidth}}
\caption{Polarization of stars in $B$ (left) and I(right) in the field of 
A2219.}
\label{psf}
\end{figure}

In order to recover the true galaxy shapes, they have to be corrected for
various observational effects. PSF anisotropy introduces a systematic
distortion in the shapes of the images of the faint galaxies, mimicing
a lensing signal. We select a sample of moderately bright stars and
quantify the PSF anisotropy as described in Hoekstra et al. (1998).
Figure~\ref{psf} shows the observed PSF anisotropy as a function of position
on the chip for both the $B$ band (left) and $I$ band (right). The sticks
indicate the direction of the major axis, and the length corresponds to the 
amplitude of the anisotropy. Following the scheme described in Kaiser et al. (1995),
and Hoekstra et al. (1998) we are able to correct the polarizations of the faint
galaxies for PSF anisotropy.

After correction for PSF anisotropy, one still needs to correct for the fact
that seeing and the instrumental PSF (now isotropic) circularize the images.
We follow Luppino \& Kaiser (1997) and Hoekstra et al. (1998) to estimate the
'pre-seeing' shear polarizability (Luppino \& Kaiser 1997). The resulting 
correction depends on both the size and the magnitude of the galaxies,
where the correction is largest for the smallest objects.

\subsection{Weak distortion}

The distortion $g$ is related to the shear $\gamma$ and the convergence $\kappa$
via $g=\gamma/(1-\kappa)$. The weak lensing distortions are small compared to
the intrinsic ellipticities of the sources. Thus we average the shapes of a 
large number of galaxies. We weight the contribution from each object using the
uncertainty in the distortion, which includes both the contribution of the 
intrinsic ellipticities of the galaxies and the shot noise 
(Hoekstra, Franx \& Kuijken 2000).

\begin{figure}
\centerline{\psfig{figure=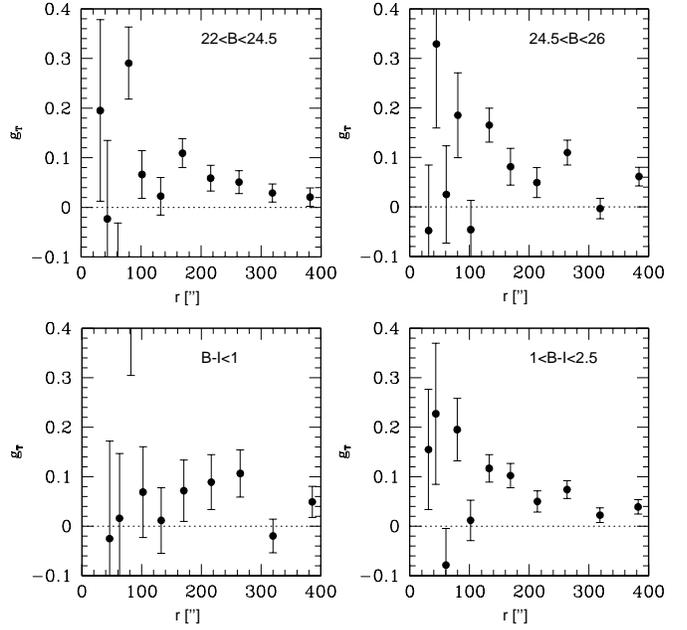,width=0.5\textwidth}}
\caption{Distortion $g_T$ for the bright, faint, blue and red objects.}
\label{g_T_mag}
\end{figure}

\begin{figure}
\centerline{\psfig{figure=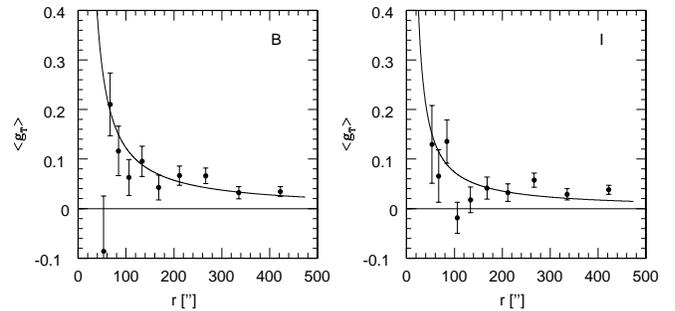,width=0.5\textwidth}}
\caption{Distortion $g_T$ and SIS fit for objects detected in the B image
($\sigma=1225\,\pm 100\,km\,s^{-1}$) and in the I image ($\sigma=975\pm 
125\,km\,s^{-1}$), excluding cluster members.}
\label{g_T_B_I}
\end{figure}

\begin{figure}
\centerline{\psfig{figure=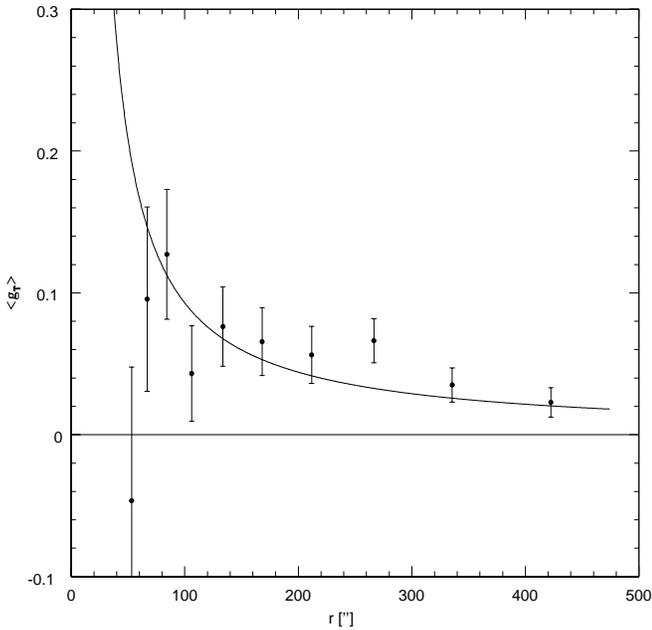,width=0.5\textwidth}}
\caption{Distortion $g_T$ for objects detected in the B image and/or I
image, excluding cluster members. The solid line is a SIS fit to the data
($\sigma=1075km\, s^{-1}$)
using the redshift distribution of figure \ref{N_z} for the background 
population.}
\label{g_T}
\end{figure}

Selecting objects according to their magnitude
reveals a higher distortion for faint objects with respect to the
bright ones (figure \ref{g_T_mag}). All objects detected in B or I
with colour $B-I$ outside the cluster elliptical sequence are
considered here, which corresponds to a number density of 50 objects
per square arcmin. 
When all objects besides cluster members are considered,
a higher signal is detected in the $B$ image than in the
$I$ image probably because of a remaining contamination by cluster members 
in $I$ (figure \ref{g_T_B_I}).
When restricting the sample to objects detected in both B and I and
excluding the cluster members, the weak distortions are clearly detected 
out to 2 $h^{-1}_{50}Mpc$ away
from the cluster center (figure \ref{g_T}). A singular isothermal sphere 
fit accounts for the data with a velocity dispersion of $1075\pm 100\, 
km\, s^{-1}$. 

The derived velocity dispersion clearly depends sensitively on the assumed 
redshift distribution of the sources. In the above estimates we 
adopted the photometric $N(z)$ derived in the central field for this purpose
(Figure \ref{N_z}).
However, this could be uncertain for a variety of reasons. First, photometric
redshifts cannot be derived or are poorly constrained for very faint
objects as the photometry reliability decreases quickly while weak
shear increases with magnitude (figure \ref{g_T_mag} for the bright
and faint samples). Second, because of gravitational lensing
magnification, the redshift distribution of background objects should
contain more high-$z$ objects in the center than in the outside region
of the cluster and photometric redshifts can only be determined in the
central field. Third, the width of the cluster peak around $z=0.225$
(figure \ref{N_z}) appears to be very broad.  
This means that many cluster members appear at $z_{phot}=0.3 -
0.4$ which bias the $N(z)$ towards low redshifts.
In this analysis all objects with $z_{phot}>3$
were removed as their redshift identification is still uncertain.

A major advance offered by the availability of multiband optical and 
near-IR data is that we can select objects in different photometric redshift 
intervals to verify the robustness of the derived mass. We have done this 
in such a way so as to maintain a reasonably-sized sample.

\begin{figure}
\centerline{\psfig{figure=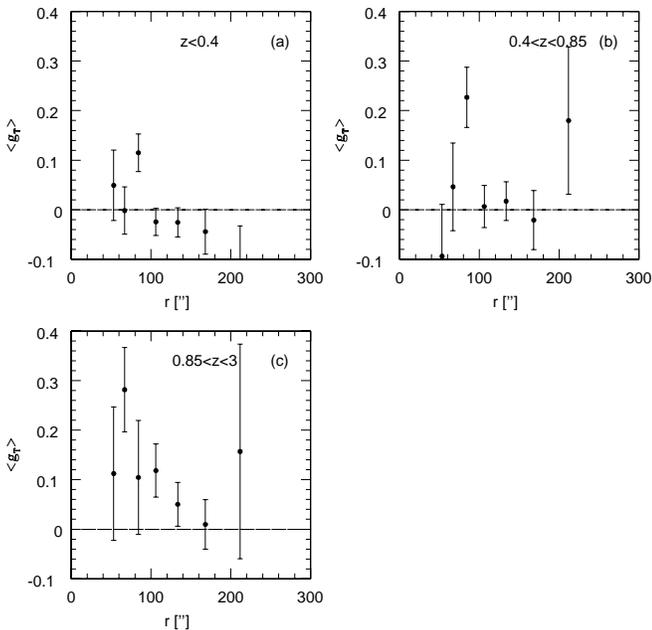,width=0.5\textwidth}}
\caption{Distortion $g_T$ for objects selected 
according to three different photometric redshift intervals: 
$z_{phot}<0.4$ (a), $0.4<z_{phot}<0.85$ (b), $0.85<z_{phot}<3$ (c).}
\label{g_T_z}
\end{figure}

Selecting objects according to their photometric redshift shows a
higher distortion when one goes to high $z$ as expected (figure \ref{g_T_z} 
a, b, and c). The number of objects involved is: 330 at $z_{phot}<0.4$, 291
at $0.4<z_{phot}<0.85$ and 279 at $0.85<z_{phot}<3$. No significant signal 
is found in the first bin (a) as expected for cluster and foreground
galaxies. The signal is marginal when the sources are selected between 
$z=0.4$ and $z=0.85$ (b) and becomes obvious in the highest redshift bin
$0.85<z<3$ (c). This is a good verification of the reliability of photometric 
redshifts for weak lensing. 

\begin{figure}
\centerline{\psfig{figure=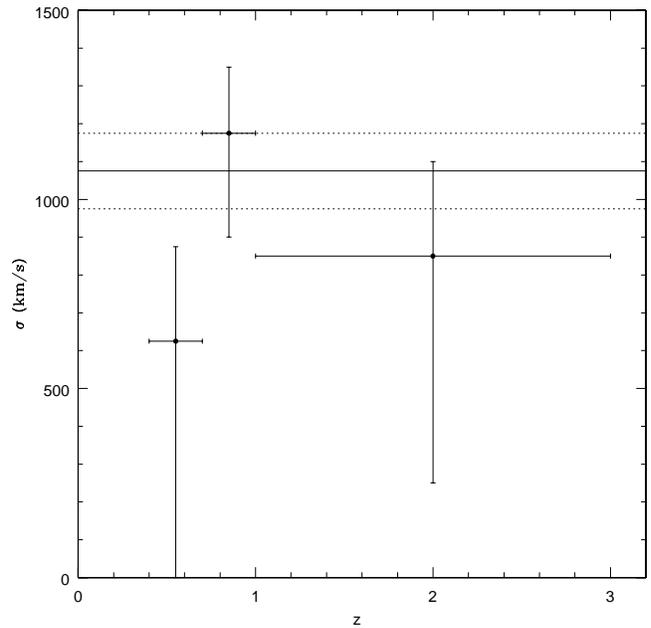,width=0.5\textwidth}}
\caption{Velocity dispersion of the lens derived in three redshift intervals 
by fitting the observed distortion profile with a SIS. The value derived with 
the whole sample ($\sigma=1075\,\pm 100 km\, s^{-1}$, figure \ref{g_T}) is 
shown as a solid horizontal line}
\label{sigma_z}
\end{figure}

In order to check whether the signal detected in each redshift bin is consistent
with the expected one given the velocity dispersion and redshift distribution 
adopted previously, 
a fit of the distortion profiles in the redshift bins [0.4,0.7], [0.7,1.0] and 
[1.0,3] has been performed assuming a SIS for the lens (figure \ref{sigma_z}).
The velocity dispersion found in the last two bins is in good agreement 
with the value derived with the 
whole sample in figure \ref{g_T} ($\sigma=1075\pm 100 km\, s^{-1}$).
The lower velocity dispersion found in the first redshift interval 
comes from the determination of photometric redshifts
which cannot lead to a cluster peak at $z=0.225$ as narrow as what a
spectroscopic survey would give. The peak in figure \ref{N_z} is much
broader than what it really is and many cluster members appear at 
higher redshifts and, as a
result, the signal at $z<0.7$ is contaminated by these unlensed objects. 
This leads to a smaller amplitude of the distortion profile and, hence, to 
a lower velocity
dispersion. On the contrary, the redshift bins [0.7,1.0] and [1.0,3] give 
coherent values which means that most of the signal detected in the
global sample (figure \ref{g_T}) comes from this redshift range.

Another way of investigating the mass distribution is given by the 
magnification bias and Gray et al. (2000) have measured the gravitational 
depletion of number counts 
in the infrared for A2219. After fitting the depletion curve by a singular 
isothermal sphere, they derived a slightly lower value for the velocity 
dispersion, $\sigma=842^{+99}_{-84} km.s^{-1}$ assuming that the sources 
lie at $z\simeq 1$.

\subsection{Mass estimates}
The observed distortion $g_T$ is related to the dimensionless 
surface mass density $\kappa$, also called convergence, via the 
parameter $\zeta$:
$$\zeta (r,r_{max})={2 \over 1-(r/r_{max})^2} \int_r^{r_{max}} g_T(r)
d \, ln(r)$$
and also
$$\zeta(r,r_{max})=\overline{\kappa_1}(r) - \overline{\kappa_2}(r,r_{max})$$
where $\kappa_1(r)$ is the average $\kappa$ inside radius $r$ and
$\kappa_2(r,r_{max})$ is the average $\kappa$ between radius $r$ and $r_{max}$.

Then, a lower limit to the mass inside radius $r$ is given by

$$M_{lim}=\pi r^2 \zeta(r) \Sigma_c$$

where $\Sigma_c={c^2 \over 4 \pi G} <{D_s \over D_l D_{ls}}>$.

\begin{figure}
\centerline{\psfig{figure=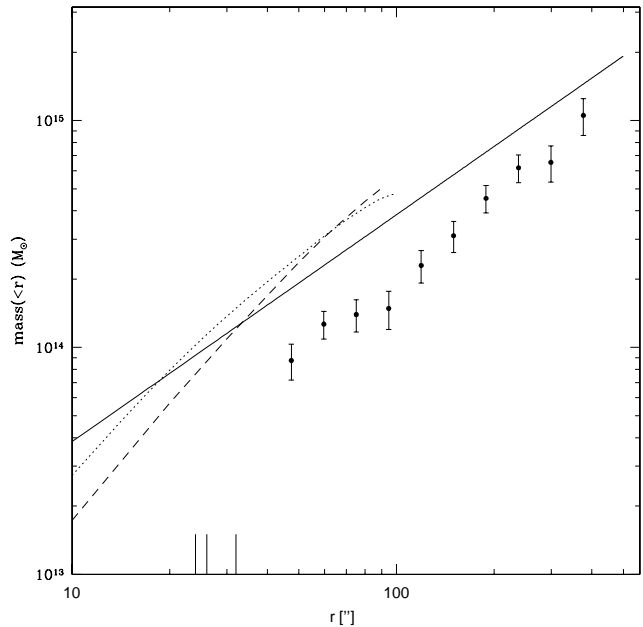,width=0.5\textwidth}}
\caption{Radial mass profile of the cluster in solar masses. Short dashed line: 
mass model
derived with strong lensing constraints following Kneib's algorithm,
dotted line: mass model derived with AbdelSalam et al. method, solid line:
mass profile corresponding to a singular isothermal sphere with a velocity
dispersion of 1075$km\, s^{-1}$. Filled
dots with error bars correspond to the lower limit on the mass profile
derived from the distortion $\zeta(r)$. The three vertical lines give
the distance of the giant arc and objects A and B with respect to the 
central cD. 100\arcsec=458$h_{50}^{-1}\,kpc$ at $z=0.225$.}
\label{mass}
\end{figure}

The photometric redshift distribution is used here to estimate
$\Sigma_c$. The resulting lower limit on the radial mass profile is
displayed in figure \ref{mass}.

The fit of the distortion profile by a singular isothermal sphere
(figure \ref{g_T}) is in good agreement with the mass profiles
coming from the two strong lensing models, at least in the inner
45\arcsec.  Moreover, these two mass models lie satisfactorily above
the lower limits derived by the $\zeta$ profile.  As $\Sigma_c$ is
very sensitive to the sources redshift distribution, the SIS profile
would be lower by 20\% if the redshift distribution of Fernandez Soto
et al. (1999) is used instead of the N(z) from figure \ref{N_z}
($\sigma=950 km\,s^{-1}$ instead of 1075$km \, s^{-1}$). This
distribution was determined in the HDF-N based on photometric
redshifts and is restricted here to $B<26$.
According to the SIS fit, the mass inside a radius of 300$h_{50}^{-1}\, kpc$ is 
$2.5\,10^{14}\,M_{\odot}$ ($4.8\, 10^{14}\,M_{\odot}$ for A370, B\'ezecourt 
et al. 1999) and $4.2\, 10^{14}\,M_{\odot}$ inside 500$h_{50}^{-1}kpc$ ($4.0\, 
10^{14}\, M_{\odot}$ in AC114, Natarajan et al. 1998).

The SIS fit provides also
an estimate of the mass to light ratio inside 1$h_{50}^{-1}\, Mpc$: $M/L=210 
h_{50}$ in the $B$ band using total magnitudes given by 
Sextractor. Cluster galaxies are selected according to their $B-I$ 
and the resulting counts per magnitude are fitted by a Schechter luminosity 
function with parameters $\alpha=-0.9$ and $M_{\star}=-20.2$. 

\begin{figure}
\centerline{\psfig{figure=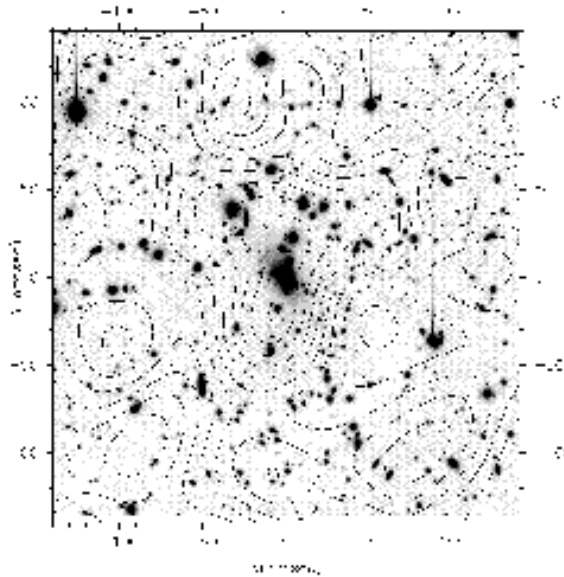,width=0.45\textwidth}}
\caption{Mass map derived from weak lensing overlayed on a $I$ band image of A2219.}
\label{mass_map}
\end{figure}

The mass map producing the observed shear field is displayed in figure
\ref{mass_map} exhibiting a peak coinciding with the central cD.  Two
extensions are also visible: one towards galaxy G2 and another one
which corresponds fairly well with a group of galaxies 45\arcsec away
to the north east of the cD.  Hence, the three mass maps obtained
using strong and weak lensing constraints agree very well with each
other. This elongated structure follows also the X-ray emission shown
in Smail et al. (1995).

The mass models in the strong and weak regime are based on
different methods and their combination indicates that the slope of 
$M(<r)$ is smaller in the cluster periphery
than in the center (figure \ref{mass}). Comparison of both
types of models in CL1358+62 (Hoekstra et al. 1998) showed that the
weak lensing analysis underestimates the mass in the cluster center
with respect to the mass derived with the constraint given by the arc
at $z=4.92$ (Franx et al. 1997).  A weak lensing analysis of A2218
gives a similar underestimate of the central mass (Squires et
al. 1996) while we get comparable values at the location of the two multiple
image systems.

\section{Conclusions}
We have presented a multicolor, wide field imaging study of A2219 and studied
the lensing properties of the cluster with it. A new feature of this analysis
is the combination of optically measured shear with photometric redshifts 
derived from $U$ through $H$ colors. 

Multicolour photometry has first contributed in
identifying a three image system at a probable redshift of 3.6 and,
according to colour differences, it has ruled out an old one. 
Deriving photometric
redshifts for the whole field is the second result of this multiband
approach. The redshift distribution of the background sources is then
accessible and shows mainly objects at $z<1.5$ with higher-$z$
candidates. This technique can be powerful only if infrared data are
available as the $H$ image taken with the CIRSI instrument fixes the
slope at large wavelengths determined by the amount of old stars.
Future developments should improve the photometric estimates of redshifts 
in order to remove ambiguities and clarify the status of $z>3$ objects. 
Arcs redshifts require spectrocopic confirmation as well.

The identification of two multiple image systems gives strong
constraints on the central mass distribution. Two modellings were
derived by different methods, one assuming that mass follows light
while the other one accepts more freedom in the location of the mass
clumps.  Both models give similar results concerning the total mass
and its increase rate. However, the $z=3.6$ system is not perfectly
well reproduced by the models, an offset in the location of the third
image or additional unseen counter images are found.

The wide field provided by the WHT Prime Focus reveals distortions of
background galaxies out to 1.5$h_{50}^{-1}\,Mpc$. The distortion profile is
consistent with a singular isothermal sphere, with a velocity
dispersion of 1075$km\,s^{-1}$ when the background galaxies redshift
distribution is assumed to be the one derived from photometric
redshifts. This value corresponds to a mass to light ratio of 210 in
the $B$ band. We show also that the lensing strength
depends on photometric redshifts in the expected way.
Considering a $N(z)$ coming from HST
observations gives a lower velocity dispersion ($\sigma=950 km\,
s^{-1}$) as more objects are present at higher redshifts, requiring
less mass to distort them.
Strong and weak lensing
observations combine to give a consistent mass model of the cluster
over the radius range $100\,h_{50}^{-1}\,kpc$ to $1.5\, h_{50}^{-1}\,Mpc$
and results in a total mass within radius 1$Mpc$ of $8.3\,10^{14}\,M_{\odot}$.

\acknowledgements This research has been conducted under the auspices
of a European TMR network programme made possible via generous
financial support from the European Commission ({\tt
http://www.ast.cam.ac.uk/IoA/lensnet/}). We are also grateful to J.P
Kneib who makes his code {\it Lenstool} available for modelling the
mass distribution of clusters lenses, Roser Pell\'o for the code {\it hyperz}
and I. Smail for the already published $U$ and $V$ images.

\end{document}